\documentclass[pre, twocolumn, superscriptaddress,showpacs, floatfix]{revtex4}

\usepackage{psfrag} \usepackage{amsmath,amsfonts,amssymb}
\usepackage{latexsym} \usepackage{graphicx}

\newcommand{\avg}[1]{\left\langle #1\right\rangle}
\newcommand{\stat}[1]{\overline{ #1}}
\newcommand{\deri}[2]{\frac{\textrm{d}#1}{\textrm{d}#2}}
\newcommand{\drho}{\delta \rho}
\newcommand{\dJ}{\delta J}
\newcommand{\Erf}{\textrm{Erf}}
\newcommand{\D}{\mathcal{D}}

\begin{document}
\title{Dynamic correlation functions and Boltzmann Langevin approach
  for a driven one dimensional lattice gas}

\author{Paolo Pierobon} \affiliation{Arnold Sommerfeld Center and
  CeNS, Department of Physics, Ludwig-Maximilians-Universit\"at
  M\"unchen, Theresienstrasse 37, D-80333 M\"unchen,
  Germany}\affiliation{Hahn-Meitner Institut, Abteilung Theorie,
  Glienicker Str.100, D-14109 Berlin, Germany}

\author{Andrea Parmeggiani}
\affiliation{Dynamique Mol\'eculaire des Interactions Membranaires,
  CNRS-UMR5539, CC107, Place Eug\'ene Bataillon, 34095 Montpellier
  Cedex 05, France} 

\author{Felix von Oppen} \affiliation {Fachbereich Physik, Freie
  Universit\"at Berlin, Arnimallee 14, D-14195 Berlin, Germany}

\author {Erwin Frey} \affiliation{Arnold Sommerfeld Center and CeNS,
  Department of Physics, Ludwig-Maximilians-Universit\"at M\"unchen,
  Theresienstrasse 37, D-80333 M\"unchen,
  Germany}\affiliation{Hahn-Meitner Institut, Abteilung Theorie,
  Glienicker Str.100, D-14109 Berlin, Germany}
  
\date{\today}

\begin{abstract} 
  We study the dynamics of the totally asymmetric exclusion process
  with open boundaries by phenomenological theories complemented by
  extensive Monte-Carlo simulations.  Upon combining domain wall
  theory with a kinetic approach known as Boltzmann-Langevin theory we
  are able to give a complete qualitative picture of the dynamics in
  the low and high density regime and at the corresponding phase
  boundary.  At the coexistence line between high and low density
  phases we observe a time scale separation between local density
  fluctuations and collective domain wall motion, which are well
  accounted for by the Boltzmann-Langevin and domain wall theory,
  respectively. We present Monte-Carlo data for the correlation
  functions and power spectra in the full parameter
  range of the model.
\end{abstract}

% 02.50.Ey Stochastic processes
% 05.60.-k Transport processes
% 05.40.-a Fluctuation phenomena, 
%          random processes, noise, and Brownian motion

\maketitle

\section{Introduction}

One-dimensional driven lattice gases are an interesting field of
non-equilibrium statistical mechanics where collective effects give
rise to unexpected non-trivial behavior such as phase transitions,
pattern formation, long-range order and anomalous diffusion. In this
class of problems the totally asymmetric simple exclusion process
(TASEP) represents an exactly solvable case which proved to model some
generic features of several systems from a rather diverse range of
fields: biological transport phenomena (e.g. ribosomes moving on mRNA
tracks \cite{macdonald-gibbs-pipkin:68} and molecular motors moving
along microtubules \cite{parmeggiani-franosch-frey:03,
lipowsky-klumpp-nieuwenhuizen:01}), traffic
\cite{chowdhury-santen-schadschneider:00}, single file diffusion
\cite{wei-bechinger-leiderer:00} and even economics
\cite{willmann-schuetz-challet:02}.

\begin{figure}[htbp]
\begin{center}
  \input{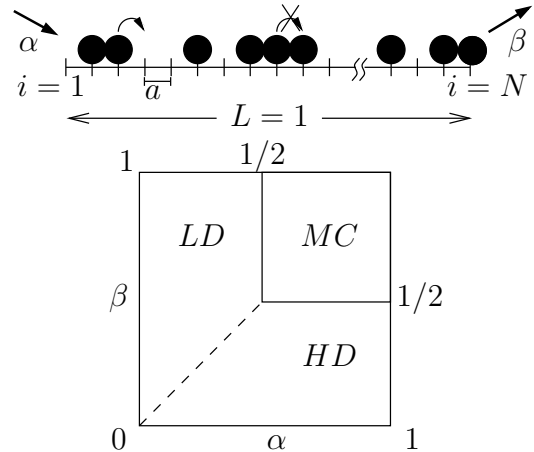}
\caption{\label{fig:model} (a) Cartoon of the TASEP model and 
  (b) phase diagram as a function of the entrance and exit rates,
  $\alpha$ and $\beta$, respectively, showing the low density (LD),
  high density (HD) and maximal current (MC) phase.}
\end{center}
\end{figure}

% Definition of the model:
In the TASEP one considers a system of identical particles moving
uni-directionally with a constant rate along a finite one-dimensional
lattice with sites labeled by $i\!=\!1,...,N$ (see
Fig.~\ref{fig:model}). The lattice spacing is $a=L/N$ with $L$ being
the total length of the system. The microscopic state of the system is
characterized by occupation numbers $n_i$ which are binary variables
with only two possible values $n_i \in \{0,1\}$, i.e. we impose a
hard-core repulsion between the particles. Of particular interest are
systems with open boundaries, where particles enter the system at the
left end with a rate $\alpha$ and leave at the right end with a rate
$\beta$. Moreover we use sequential dynamics, appropriate for
biological systems, where each particle moves according to an
``internal clock''; a parallel update would be more realistic for
vehicular traffic.

% Short review of the phase diagram and the density profile:
The stationary state (current and density profiles) of this driven
lattice gas model has been studied in great detail over the last years
(for recent reviews see e.g. Refs.~\cite{derrida:98, b:schuetz:01}).
In open systems it has been found by some exact
methods~\cite{derrida-domany-mukamel:92, schuetz-domany:93,
  derrida-etal:93} that there are three different phases: a high
density, a low density and a maximal current phase;
compare Fig.~\ref{fig:model}. The coexistence line $\alpha = \beta <
1/2$ marks a discontinuous transition between the high and low density
phase, while along the lines $\alpha = 1/2 \leq \beta$ and $\beta =
1/2 \leq \alpha$ the transitions are continuous. There are interesting
correlation effects reflected in the shape of the density profiles at
the boundaries. It is only along the disorder line, $\alpha + \beta =
1$, where the density profile is flat.

% Review of the state of the art in dynamics
The dynamic properties of the TASEP model are much less studied. For
closed systems exact analytical results~\cite{dhar:87, gwa-spohn:92,
  derrida-evans-mukamel:93, derrida-lebowitz-speer:01} have been
derived for the largest relaxation time $\tau$. It is found that
$\tau$ diverges with system size as $\tau \propto L^z$ with a dynamic
exponent of $z=3/2$.  For open systems no exact results are available.
The spectrum of relaxation times has been studied for small systems
using exact enumeration techniques~\cite{bilstein-wehefritz:97,
  dudzinsky-schuetz:00}. More recently density matrix renormalization
group studies~\cite{nagy-appert-santen:02} have shown that the largest
relaxation times are finite unless $\alpha = \beta = 1/2$, where $\tau
\propto L^{3/2}$ as for the periodic system; the results also indicate
that $z=3/2$ in the whole maximal current phase. These studies also
confirm the results of a phenomenological approach known as domain
wall theory~\cite{kolomeisky-etal:98, dudzinsky-schuetz:00}.  In this
coarse-grained description of the dynamics it is assumed that each
particle reservoir at the system's boundary independently fixes a
density. The two domains are then joined in the bulk by a phase
boundary (domain wall), which performs a random walk due to the
randomness of particle flux at the boundaries; the domain wall (DW)
moves left whenever a particle enters the system and right whenever a
particle exits the system. Along the coexistence line the DW
theory accounts even for the power spectrum at small frequencies (long
time)~\cite{takesue-mitsudo-hayakawa:03}.

DW theory does not describe the dynamics of local density
fluctuations. This is evident for periodic systems where the average
density profile is flat. In this case it has been shown that the
dynamics of the TASEP model maps onto the noisy Burgers
equation~\cite{lebowitz-presutti-spohn:88}. It is actually this
mapping which allows for the implementation of some exact methods such
as the Bethe ansatz~\cite{gwa-spohn:92}. Further progress has also
been made upon using mode coupling~\cite{vanberejen-kutner-spohn:85}
and renormalization group~\cite{forster-nelson-stephen:77} theories.

This correspondence between the lattice gas model and a Langevin
equation for the periodic case suggests to look for a similar mapping
for open systems. In this context, it is useful to recall a successful
method in quantum many-body systems, known as Boltzmann-Langevin
(BL) equations.  This approach which was first introduced to describe
electron transport in the presence of disorder and phonon scattering
\cite{kogan-schulman:69, b:kogan:96}, describes not only the average
of the electronic distribution function, but also its fluctuations.
This is achieved by amending the Boltzmann kinetic equation with a
Langevin source which takes the stochastic nature of collisions into
account. In the context of electronic transport through
nanostructures, this method has recently been widely employed in
studies of shot noise (for a review see \cite{blanter-buettiker:00}). A
close analogy with the lattice-gas model studied in this paper arises
for the following reason: the requirement of no double occupation of
sites has its direct analog in the Pauli principle which forbids
double occupation of electronic states.

The outline of the paper is the following. In the following section we
will discuss how the Boltzmann-Langevin approach can be applied for
the TASEP model and calculate the correlation function of the
linearized version. Interestingly, it will turn out that the resulting
Boltzmann-Langevin equation reduces to the noisy Burgers equation for
$\alpha = \beta = 1/2$. In section \ref{sec:mc_method} we give a short
description of the Monte-Carlo methods used to analyze the correlation
functions of the TASEP model.  Results obtained from these simulations
are compared with the analytical results from the Boltzmann-Langevin
and the DW theories in section \ref{sec:mc_results}. Finally, we
present a short summary and some outlook. In Appendix
\ref{sec:criticality} we give a thorough discussion of the behavior of
the correlation function right at the critical point.

\section{Stochastic equations of motion}
At a given time $t$ the microscopic state of the system is
characterized in terms of occupation numbers $\{ n_i =0, 1\}$. The
dynamics, that we described as a set of rules in the previous
section, can be formulated in terms of a {\em quantum Hamiltonian
  representation}~\cite{doi:76, grassberger-scheunert:80, peliti:84}.
In the bulk the corresponding Heisenberg equations for the occupation
number operators $n_i(t)$ have the form of a lattice continuity
equation
\begin{subequations}
\label{eq:micro}
\begin{eqnarray}
\label{eq:micro_conservation}
 \partial_t n_i(t) =  J_{i-1} (t) -  J_{i} (t) 
\end{eqnarray}
with the current operator
\begin{eqnarray}
\label{eq:micro_current}
 J_i (t) = n_i(t) \, \left( 1- n_{i+1}(t) \right) \, .
\end{eqnarray}
\end{subequations}
The effect of the entrance and exit rates is equivalent to constant
particle reservoirs of density $\alpha$ and $1-\beta$ at auxiliary
sites $i=0$ and $i=N+1$, respectively.

There are several levels of approximation in dealing with the dynamics
of the system. If correlation effects are neglected altogether one
arrives at a set of rate equations for the average particle density
$\rho_i (t) = \avg {n_i (t)}$, which have a form identical to
Eq.~(\ref{eq:micro}) with $n_i$ replaced by $\rho_i$. To arrive at these
equations one has to take the average of Eq.~(\ref{eq:micro}) and
neglect correlations in the spirit of a mean-field or a random phase
approximation
\begin{equation}
\avg{n_i(t) \, n_{i+1}(t)} 
\rightarrow \avg{n_i(t)} \, \avg{n_{i+1}(t)}  \, .
\end{equation}
Then, in the stationary limit the rate equations are equivalent to a
nonlinear map 
\begin{eqnarray}
  \label{eq:nonlinear_map}
  \stat{\rho}_i (1-\stat{\rho}_{i+1}) = J
\end{eqnarray}
with a constant stationary current $J$. Upon exploiting the properties
of this map one can easily reproduce the full phase diagram of the
TASEP~\cite{macdonald-gibbs-pipkin:68}. Actually, it turns out that
the phase diagram~\cite{derrida-domany-mukamel:92} obtained in this
way is identical to the one obtained from an exact solution of the
TASEP in the stationary
limit~\cite{derrida-etal:93,schuetz-domany:93}. The density profiles
obtained from such a mean-field approach miss correlation effects,
especially in the maximal current phase, and the fluctuations of the
domain walls.

\subsection{The Boltzmann-Langevin approach}
To go beyond rate equations we follow a line of arguments which leads
to what is known as Boltzmann-Langevin (BL) equations in studies of
non-equilibrium transport in electron
systems~\cite{kogan-schulman:69,b:kogan:96}.  The right hand side of
Eq.~(\ref{eq:micro_conservation}) has a form similar to the
\emph{collision integral for impurity scattering} in the Boltzmann equation balancing in-going and
out-going currents. In order to account for fluctuation effects around
the stationary state, we express both the current and the density as
the sum of a deterministic and a fluctuating part
\begin{subequations}
\label{eq:expansion}
\begin{eqnarray}
  n_i &\approx& \stat{\rho}_i + \drho_i \equiv \rho_i \, , \\
  J_i &\approx&  \rho_i (1- \rho_{i+1}) + \dJ_i \, .
\end{eqnarray}
\end{subequations}
Since we will use the BL approach only for those
regions in the phase diagram where the stationary density is to a good
approximation spatially constant we may set $\stat{\rho}_i =
\stat{\rho}$. This applies for both the high and low density phase,
but not for the phase boundary $\alpha = \beta \leq \frac12$, where in
addition to density fluctuations on small scales we also have domain
wall motion on large scales. The latter modes are obviously not
accounted for in the BL formulation. One also has to
be cautious in the maximal current phase where boundary layer profiles
decay only algebraically as one moves from the boundaries towards the
bulk~\cite{derrida-domany-mukamel:92}.

Upon inserting Eq.~(\ref{eq:expansion}) into the equations of motion,
Eq.~(\ref{eq:micro}), we find a coupled set of Langevin equations for the
density fluctuations at each site of the lattice
\begin{eqnarray}
\label{eq:langevin_discrete}
\partial_t \drho_i(t)
&=& (1-\stat{\rho}) 
    \left[ \drho_{i-1}-\drho_i \right] - \stat{\rho}
    \left[\drho_i-\drho_{i+1} \right] \nonumber\\
&&+ \drho_i( \drho_{i+1}-\drho_{i-1}) -(\dJ_i-\dJ_{i-1}) \, .
\end{eqnarray}
In order to close these equations we still need to specify the current
fluctuations $\dJ_i$. This can be done by exploiting the fact that the
occupation numbers are binary variables, which immediately implies
that $J_i^2 = J_i$ \footnote{Note that our derivation {\it differs}
  from the argument used in the conventional Boltzmann-Langevin
  theory. The latter would only yield the low-current approximation
  $\rm Var[J_i]=\langle J_i\rangle$.}.  Hence the variance of the current
at a particular site is given by
$\text{Var}[J_i]=\avg{J_i}(1-\avg{J_i})$.  To be consistent with the
approximations already made, we set $\avg{J_i}\approx \stat{\rho} (1 -
\stat{\rho})$ and finally get
\begin{eqnarray}
  \label{eq:noise_amplitude}
  \text{Var} [J] =\stat{\rho} (1 - \stat{\rho}) 
                     (1- \stat{\rho} (1 - \stat{\rho})) \, .
\end{eqnarray}
Our final assumption is that correlations in the current fluctuations
are short ranged in space and time such that we can write
\begin{eqnarray}
  \label{eq:noise_correlation}
  \avg{\dJ_i (t) \dJ_j (t')} 
  = \text{Var} [J] \, \delta_{ij} \,\delta (t-t') \, .
\end{eqnarray}
Note that local current fluctuations are due to the fact that each
particle advances randomly at a given rate (set equal to $1$), with an
exponential distribution of waiting times (in the low density limit).

\subsection{Gradient expansion}
We will now derive a continuous version of the discrete
BL equations, Eq.~(\ref{eq:langevin_discrete}). To this
end we set $x=i a$ and introduce fields $\phi (x,t) = \drho_i (t)$ and
$\eta (x,t) = \dJ_i (t)$ for the density and current fluctuations,
respectively. Then we get to leading order in a gradient expansion
\begin{eqnarray}
\label{eq:langevin}
\partial_t \phi (x,t) + (v - 2 \phi) \partial_x \phi 
= \frac12  \partial_x^2 \phi 
  -  \partial_x \eta  \, ,
\end{eqnarray}
where from now on we measure all length scales in units of the lattice
spacing $a$. Eq.~(\ref{eq:langevin}) has previously been derived along
similar lines in Ref.~\cite{krug:91}. The noise correlations are given
by
\begin{eqnarray}
  \avg{\eta(x,t) \eta(x',t')} = A \delta(x-x')\delta(t-t')
\end{eqnarray}
with an amplitude $A= \stat{\rho} (1 - \stat{\rho}) [1- \stat{\rho} (1
- \stat{\rho}) ]$. We have also introduced the collective velocity
$v=1-2\stat{\rho}$, which happens to coincide with the expression
obtained from the exact non-equilibrium fluctuation-dissipation theorem
$v = \partial_\rho J(\rho)$ of an infinite lattice
gas~\cite{kolomeisky-etal:98}.  Note that $v$ changes sign at
$\stat{\rho}=\frac12$ where the stationary current becomes maximal.

The convective nonlinearity $\phi \partial_x \phi$ in
Eq.~(\ref{eq:langevin}) can be read as a ``shift'' in the collective
velocity due to fluctuations, which we expect to become important for
small $v$, i.e. close to the phase boundaries between the low and high
density phases and the maximal current phase.  For densities far away
from $\stat{\rho}=1/2$ we will neglect those nonlinearities. Then, as
will be discussed in the next subsection, one can work out all the
correlation functions explicitly. These will then be used as a
guidance for the discussion of the Monte-Carlo results in
section~\ref{sec:mc_results}.

For $\stat{\rho}=\frac12$ Eq.~(\ref{eq:langevin}) is identical to the
one-dimensional {\em Burgers equation}
\cite{forster-nelson-stephen:77}, which can be mapped onto the
Kardar-Parisi-Zhang (KPZ) \cite{kardar-parisi-zhang:86} equation upon
introducing a new field $h$ via $\phi = \partial_x h$. The Burgers
equation is known to give the following scaling form for the
correlation function $C(x,t) = \avg{\phi(x,t) \phi(0,0)}$
\cite{forster-nelson-stephen:77},
\begin{equation}
\label{eq:scal_kpz}
   C (x,t) = x^{2\chi-2} F(t/x^z) \, .
\end{equation}
Here the roughness exponent $\chi$ describes the scaling of the width
of the interface, and the dynamic exponent $z$ characterizes the
spread in time of disturbances on the surface.  In the present case
the roughness and the dynamic exponent are known to be
\cite{forster-nelson-stephen:77}
\begin{subequations}
\begin{eqnarray}
  \chi = \frac12 \, , \quad \text{and} \quad 
  z   = \frac32 \, .
\end{eqnarray}
\label{eq:exponents}
\end{subequations}
This basically means that the system at the critical point relaxes
superdiffusively i.e. $C(0,t)\sim t^{-1/z}$ with $z<2$.

\subsection{Correlation functions of the linearized Boltzmann-Langevin
  equation} 
The linearized Boltzmann-Langevin equation is most
conveniently analyzed in Fourier space, where it reads
\begin{eqnarray}
\label{eq:BL_fourier}
 \left[ i\omega-i v q  
       + \frac12 {q^2} \right]\, \phi (q,\omega) 
 = i  q \, \eta (q,\omega) \, .
\end{eqnarray}
{F}rom this one can immediately infer for the correlation function
$C(x-x',t-t') = \avg{ \phi (x,t) \phi(x',t') }$ in Fourier space
\begin{eqnarray}
\label{eq:corrfou}
C(q,\omega) 
= \frac{A q^2}  {\left(\omega- vq \right)^2+ \frac14 q^4} \, ,
\end{eqnarray}
and direct space
\begin{eqnarray}
\label{eq:corrreal}
C(x,t) 
= \frac{A}{\sqrt{2\pi \left|t\right|}}
  \exp \left[-\frac{(x-vt)^2}{2|t|}\right] \, .
\end{eqnarray}
Note that $C(x,t)$ is a Gaussian whose center moves with a drift
velocity $v=1-2\stat{\rho}$ and which broadens diffusively starting
from a $\delta$-function at $t=0$; height $H(t)$ and width $W(t)$ are
given by
\begin{align}
  H(t)=\frac{A}{\sqrt{2\pi |t|}} \, ,\\
  W(t)=2\sqrt{|t|} \, .
\end{align}
The on-site correlation function decays exponentially for $ v \neq 0$,
\begin{eqnarray}
 C(0,t)=\frac{A}{\sqrt{2\pi \left|t\right|}}
\exp\left[-\frac{v^2}{2} \left|t\right|\right] \, .
\label{eq:corr_relax}
\end{eqnarray}
For $v=0$, $C(0,t)$ scales as $t^{-1/2}$ as well as $C(0,\omega)$ like
$\omega^{-1/2}$.  Note also that the static limit of the correlation
function for the linearized theory is
\begin{equation}
\label{eq:statlim}
\lim_{\omega\to 0}C(q,\omega)=\frac{4A}{4v^2+q^2}
\end{equation}
which is identical to the correlation function for a Landau theory in
a Gaussian approximation usually found in equilibrium thermodynamics
\cite{b:goldenfeld:92}. This result suggests that linearized BL theory
can be viewed as the analog of the Gaussian approximation for driven
lattice gases.  The form of Eq.~(\ref{eq:statlim}) implies a
correlation length of $1/2v$ which diverges at the critical point
$\overline\rho=1/2$.

\section{Monte-Carlo simulation methods}
\label{sec:mc_method}

Of course, linearized Boltzmann-Langevin theory is valid only for very
low densities (i.e. low values of $\alpha$). To go beyond this low
density limit and test the range of validity of the linearized BL
approach we have performed extensive Monte-Carlo (MC) simulations.  To
this end we have chosen the random sequential updating algorithm by
Bortz, Kalos and Lebowitz (BKL- or $n$-fold
method)~\cite{bortz-kalos-lebowitz:75, b:landau-binder:00}.  Since it
keeps a list of all sites which are possible candidates for a
successful update it is (for the present case) faster than
conventional methods. Moreover it constitutes a reliable way to
simulate \emph{real time} dynamics and achieve an excellent quality in
terms of data and computational efficiency in both short and long time
regimes.

In a first step one generates a random number $X\in[0,1)$, that
determines which one of the following moves is chosen: a particle
entering the system, a particle leaving the system and particle at
site $i$ jumping to the right. Then, for a given move a time interval
$\Delta t$ is chosen from an exponential waiting time distribution,
where the decay time depends on the size of the list.

In all of our Monte-Carlo runs we started from a configuration
generated according to the steady state distribution in order to
reduce initial transient effects. After equilibration correlation
functions were measured and moving time averages over $O(10^7)$ time
windows were performed. Average profiles and correlations do not show
any differences between moving time and ensemble average giving
explicit proof for ergodicity of the system.

\section{Dynamic correlation functions}
\label{sec:mc_results} 
In this section we analyze the correlation function $C(x,t)$ of the
density fluctuations $\phi(x,t)$ in space and time. From now on we
take $x=0$ as the central site of the system such that the system is
confined to the interval $[-L/2,L/2]$. This is meant to minimize, at
least for short times, the influence of the system boundaries. We
assume that the system is in a stationary state at the reference time
$t=0$.

As expected from the linearized BL equation, the
simulations show that the correlation function starts from a
$\delta$-function peaked at the reference site and then moves to the
right with velocity $v$ and spreads diffusively as time progresses;
for negative times it moves to the left (see
Figs.~\ref{fig:time_series_c}). Note that our simulations confirm that
the peak of the correlation function moves with exactly the collective
velocity $v= 1 - 2 \stat{\rho}$.

In the following we are going to discuss the form and the time
evolution of the correlation function in the $\alpha$-$\beta$ plane.
Our simulations have been performed mainly along the anti-diagonal
$\alpha=1-\beta$. This has the advantage that the exact steady state
profile is perfectly flat, such that boundary effects are greatly
reduced; moreover mean field theory is predicts exactly 
$\stat{\rho}=\alpha$.

Before we enter the discussion let us have a closer look at the
characteristic time scales of the system. Correlation functions decay
on a scale (see Eq.~\ref{eq:corr_relax})
\begin{equation}
\tau_\text{relax} = 2/v^2 \, ,
\end{equation} 
but at the same time the maximum of the correlation function moves
with a velocity $v$ such that it propagates the finite length $L$ of
the track in a time
\begin{equation}
\tau_\text{prop}= \frac{L/2}{v} \, .
\end{equation}
This implies that one can observe the relaxation of the correlation
functions only if $\tau_\text{prop}/\tau_\text{relax} = v \frac L 4
\geq 1$, i.e. for rates $\alpha$ and $\beta$ not too close to the
phase boundary to the maximal current phase and of course for large
enough systems.

\subsection{Low density and high density phase}
Due to particle-hole symmetry we restrict our discussion to the low
density regime. The results for the high density phase are obtained
upon simply replacing $\alpha$ by $1-\beta$.

%\newpage
\begin{figure}[tbp!]
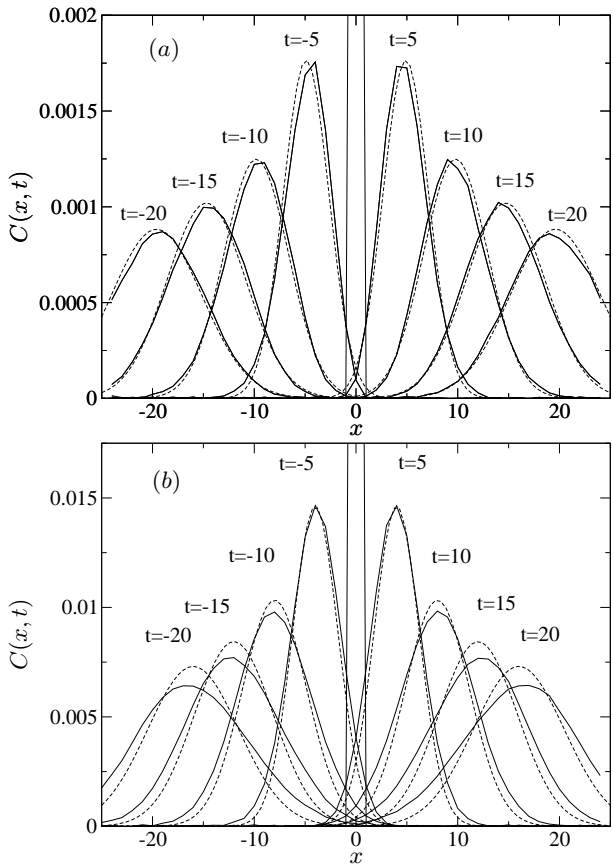

  \begin{center}
  \psfrag{C(x,t)}[][][1]{$C(x,t)$}
  \psfrag{Site}[][][1]{$x$}
  \psfrag{(a)}[][][1]{$(a)$}
  \includegraphics*[keepaspectratio, width=8cm]{a01b99N50.eps}
  \vfill
  \psfrag{C(x,t)}[][][1]{$C(x,t)$}
  \psfrag{Site}[][][1]{$x$}
  \psfrag{(b)}[][][1]{$(b)$}
  \includegraphics*[keepaspectratio, width=8cm]{a1b9N50.eps}
    \caption{\label{fig:time_series_c}
      Time and space dependent correlation function for a system of
      $50$ sites with rates (a) $\alpha=0.01$ and $\beta=0.99$ and (b)
      $\alpha=0.1$ and $\beta=0.9$.  Averages are taken over $10$
      million samples. For the diluted system there is good agreement
      between the linearized Boltzmann-Langevin theory (dashed line)
      and the Monte-Carlo simulations (solid line). For denser systems
      the agreement is only qualitative.}
  \end{center}
\end{figure}

Figures~\ref{fig:time_series_c} show time series of the correlation
function for entrance rates $\alpha=0.01$ and $\alpha = 0.1$,
respectively.  In both cases the correlation functions start from a
$\delta$-function peak at $t=0$ which then broadens diffusively with
the width scaling as $t^{1/2}$ and the height decreasing as
$t^{-1/2}$. Correspondingly the maximum propagates to the right or
left with velocity $v=1-2\stat{\rho}$ for $\stat{\rho}<1/2$ and
$\stat{\rho}>1/2$, respectively (by particle-hole symmetry).  As noted
above this is an exact result valid for all values of the entrance and
exit rates. In Fig.~\ref{fig:time_series_c} we have also shown
results for negative time to highlight the symmetry $x\to -x$ and $t\to
-t$ which appears in Eq.~(\ref{eq:corrreal}).

\begin{figure}[tbp!]
\begin{center}
\psfrag{o2}[][][1]{$t^{-1/2}$}
\psfrag{o3}[][][1]{$t^{-2/3}$}
\psfrag{0.5}[][][1]{$\alpha=0.5$}
\psfrag{0.1}[][][1]{$\alpha=0.1$}
\psfrag{0.05}[][][1]{$\alpha=0.05$}
\psfrag{0.005}[][][1]{$\alpha=0.005$}
\psfrag{Peak}[][][1]{$H(t)$}
\psfrag{Time}[][][1]{$t$}
\includegraphics*[keepaspectratio, width=8cm]{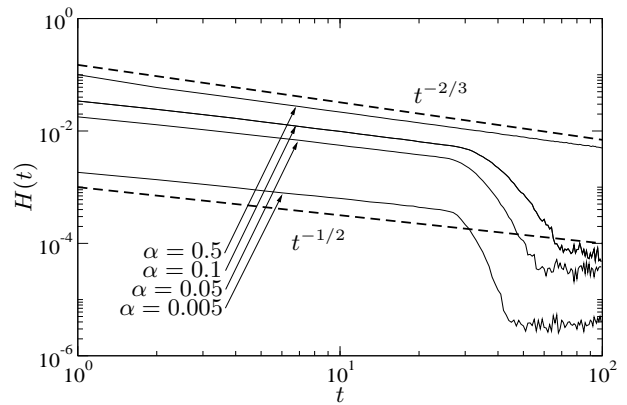}
\caption{\label{fig:peak}Peak height of the correlation function for a system 
  of $N=50$ sites and a series of entrance rates $\alpha=$ $0.005$,
  $0.05$, $0.1$, $0.5$; we have taken $\beta = 1 -\alpha$. Averages of
  the MC data are taken over $10$ millions samples. Upon approaching
  the critical point $(\alpha,\beta) = (1/2,1/2)$ the peak height
  shows a power law behavior with an effective exponent slowly
  changing from $1/2$ to $2/3$, shown as dashed lines in the graph.}
\end{center}
\end{figure}

For low values of $\alpha$, which corresponds to the low density
limit, the results from the linearized BL equations
explain the Monte-Carlo results quantitatively. The theory still gives
the correct qualitative picture for larger values of $\alpha$ but
shows significant quantitative deviations. The actual shapes of the
correlation functions have a lower height and are broader than the
linearized theory predicts. Nevertheless, the peak height measured in
the simulations still shows the $t^{-1/2}$ scaling of the linear
theory for values of $\alpha$ not to close to $\alpha = \frac12$; see
Fig.~\ref{fig:peak}.

%\newpage

As can be inferred from Fig.~\ref{fig:peak} the effective exponent
describing the peak relaxation slowly crosses over from $1/2$ to $2/3$
upon approaching the critical point $(\alpha, \beta) = (1/2,1/2)$
along the anti-diagonal of the phase diagram. The exponent $2/3$ is
identical to the inverse of the dynamic exponent $1/z$ of the
non-linear BL equation (Burgers equation); see
Eqs.~(\ref{eq:langevin}) and (\ref{eq:exponents}).

\subsection{Coexistence line ($\alpha=\beta<1/2$)}
\label{sub:coex}

At the coexistence line the density profile is characterized by a
fluctuating domain wall separating a low density from a high density
phase. The dynamics of the domain wall can be described as a symmetric
random walk with reflecting boundary
conditions~\cite{kolomeisky-etal:98, b:schuetz:01}. In addition to the
domain wall motion, a collective mode, there are still stochastic
fluctuations of the density in the low and high density wings of the
domain wall. Both of these modes should be visible in a measurement of
the density-density correlation function.

\begin{figure}[tbp!]
\begin{center}
\psfrag{C(x,t)}[][][1]{$C(x,t)$}
\psfrag{Site}[][][1]{$x$}
\includegraphics*[keepaspectratio, width=8cm]{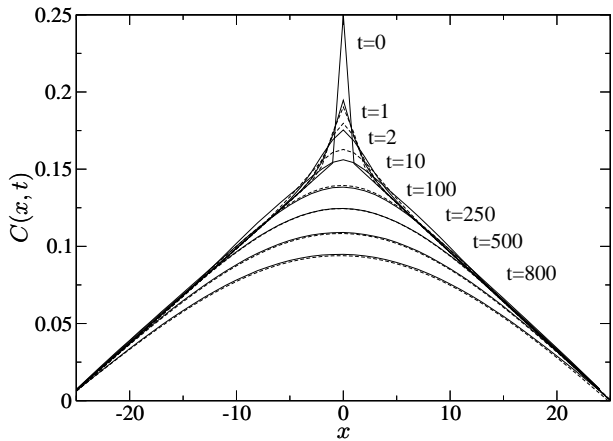}
\caption{\label{fig:alpha_0.1_beta_0.1} Time series of the correlation
  function $C(x,t)$ versus $x$ for $\alpha=0.1$ and $\beta=0.1$; the
  times are indicated in the graph. The MC data for a system with
  $N=50$ sites (solid lines) are compared with a hybrid theory
  combining local density fluctuations described by the
  Boltzmann-Langevin equation and the collective domain wall motion
  (dashed lines). Averages in the MC data are taken over $10$ million
  samples.}
\end{center}
\end{figure}

Indeed, the profile of the correlation function shows two distinct
features; see Fig.~\ref{fig:alpha_0.1_beta_0.1}. At $t=0$ there is a
sharp triangle on top of a much broader triangular base.  The sharp
tip is a result of the local density (BL) fluctuations and can be
explained as follows.  Consider the correlation function of local
density fluctuations $C^{\rm BL}_{ij} (t) = \avg { \phi_i (t) \phi_j
  (0) }$ on a lattice.  Since there are no correlations in the bulk of
the system for $t=0$ it reduces to $C^{\rm BL}_{ij} (0) = (\avg{n_i^2} -
\avg{n_i}^2) \delta_{ij}$.  Finally, upon using that the
occupation numbers are binary variables and the average density in the
middle of the system is $\avg{n_0} = \frac12$, we obtain $C^{\rm
  BL}_{i0} = \frac14 \delta_{i0}$.  The width of the sharp tip is
actually a finite size effect resulting from the linear interpolation
of the data points. The broader triangular base is explained below in the 
context of DW theory.

The MC simulations show that the sharp tip quickly relaxes and
broadens, whereas the shape of the triangular base evolves on a much
larger time scale.

%\newpage

One can rationalize this behavior upon combining results from the
DW and BL theory. We start with a discussion
of the local density fluctuations. One can derive the short time
dynamics for the correlation function from the lattice version of the
BL equation, Eq.~(\ref{eq:langevin_discrete}), imposing
the initial condition $C^{\rm BL}_{i0} (t=0) = \frac14 \delta_{i0}$.
As explicitly shown in Appendix~\ref{app:bl_lattice} one finds for the
on-site correlation function $C^{\rm BL}_{00} (t) \sim\frac1 4 -\frac
1 4 |t|+O(t^2)$, while the nearest and next-nearest neighbor
correlation functions read as $C^{\rm BL}_{01}(t)\sim t$ and $C^{\rm
  BL}_{02}(t)\sim t^2$. This explains the fast relaxation of the
central peak.

In order to understand the broadening of the triangular base we first
have to recapitulate some key results of the DW
theory~\cite{kolomeisky-etal:98, dudzinsky-schuetz:00}.  Since one can
model the domain wall as a symmetric random walker with reflecting
boundary conditions at both ends of the system, the conditional
probability of finding the domain wall at site $\xi_t$ at time $t$
given that it was at site $\xi_0$ at time $t=0$
reads~\cite{schwarz-poland:74}
\begin{align}
  P(\xi_t | \xi_0) =& \frac 1 L+\frac 2 L\sum_{i:\textrm{even}}
  e^{-\lambda_i^2\D t} \cos\lambda_i \xi_t \cos\lambda_i \xi_0\nonumber\\
  &+\frac 2 L\sum_{i:\textrm{odd}}e^{-\lambda_i^2\D t}\sin\lambda_i
  \xi_t \sin\lambda_i \xi_0 \, ,
\label{eq:propagator}
\end{align}
where $\lambda_i=i\pi/L$ and $\D = \alpha(1-\alpha) / (1-2\alpha)$ is
the diffusion constant. Note that this diffusion coefficient is
smaller than the one of the BL fluctuations (which is $1$),
$\D\sim\alpha<1$, which explains the time-scale separation mentioned
above. Averages of an observable $\mathcal O$ are understood as integrals over
the random variable $\xi_t$
\begin{eqnarray}
  \avg {\mathcal O(t,t')}= \int \! d \xi_t \int \! d \xi_{t'}
  P_{\rm st}(\xi_{t'})  \mathcal O(t,t') P(\xi_t | \xi_{t'}) \, ,
\end{eqnarray}
where $P_{\rm st}(\xi)$ is the stationary probability distribution
function. In the present case it is simply a constant, $P_{\rm
  st}=1/L$.

If one approximates the density profile of the domain wall (DW) by a
step function, $\psi (x,t) = \alpha+(1-2\alpha) \theta(x-\xi_t)$, the
correlation function can easily be calculated as
\begin{widetext}
\begin{eqnarray}
\label{eq:dw_sum}
\lefteqn{C^{\rm DW} (x,x',t-t')
  =\avg{\psi_{\rm d}(x,t) \psi_{\rm d}(x',t')} 
  -\avg{\psi_{\rm d}(x,t)}\avg{\psi_{\rm d}(x',t')}=}
  \nonumber\\
& \frac{2(1-2\alpha)^2}{L^2}\left[ \sum_{i:\textrm{even}}
\frac{e^{-\lambda_i^2\D |t-t'|}}{\lambda_i^2}
\cos\lambda_i x\cos\lambda_i x'+
 \sum_{i:\textrm{odd}} 
\frac{e^{-\lambda_i^2\D |t-t'|}}{\lambda_i^2}
\sin\lambda_i x\sin\lambda_i x'\right]\, .
\end{eqnarray}
\end{widetext}

Here we are mainly interested in the dynamics at time scales $t<
L^2/\D$ (so that the system size is large, $L^2>t\D$, and $\lambda_i$ is
infinitesimal), where the domain wall has not explored the full system
yet.  Then the sum in Eq.~(\ref{eq:dw_sum}) can be approximated by an
integral, and one finds
\begin{widetext}
\begin{eqnarray}
\label{eq:coex_DW}
C^{\rm DW} (x,t)& =&  \frac{1}{2L} \, (1-2\alpha)^2 
\left\{  
  \left[|x+L| \Erf\left(\frac{|x+L|}{\sqrt{4\D|t|}}\right)
- |x|\Erf\left(\frac{|x|}{\sqrt{4\D|t|}}\right)\right]+\right.\nonumber\\
&&+ \left.\left(e^{\frac{-{\left(x+L\right)}^2}{4\D|t|}}
- e^{\frac{-x^2}{4\D|t|}}\right)\sqrt{\frac{4\D|t|}{\pi}}
- \left(x+\frac L2\right)
\right\} \, .
\end{eqnarray}
\end{widetext}
In the limit $t\to 0$ this exactly reduces to the profile of the broad
triangular base in Fig.\ref{fig:alpha_0.1_beta_0.1}. If one would be
allowed to just sum the correlation functions obtained from domain
wall and local density fluctuations, this would fully explain the
initial shape of the correlation function. Of course, this is not
valid rigorously but seems to be a reasonable approximation. One may
argue that the validity of the approximation is due to the time and
length scale separation between the local density fluctuations and the
collective domain wall motion.

In this spirit we assume that the total density fluctuations $\Phi$
can be written as a superposition of local density and domain wall
fluctuations, $\Phi(x,t) = \phi (x,t) + \psi (x,t)$, and that these
fluctuations are uncorrelated, $\avg{\phi \psi }= \avg{\phi}
\avg{\psi}$. Then the full correlation function can be written as a
sum
\begin{eqnarray}
 C(x,t) = C^{\rm BL} (x,t) + C^{\rm DW} (x,t) \, 
\end{eqnarray}
with $C^{\rm DW} (x,t)$ given by Eq.~(\ref{eq:coex_DW}) and the local
density correlations $C^{\rm BL} (x,t)$ are obtained either from the
continuous or the lattice BL equations depending on
the time scale. Note that the density fluctuations on both wings of
the domain wall are the same since the average low ($\rho_- = \alpha$)
and high ($\rho_+= 1-\alpha$) density lead to the same noise amplitude
$A$. Hence we may describe these local density fluctuations by a
BL equation with $v=0$ and $A= \alpha (1-\alpha) (1-
\alpha (1-\alpha))$. As can be inferred from
Fig.~\ref{fig:alpha_0.1_beta_0.1} the corresponding analytical results
compare reasonably well with MC data.

\begin{figure}[tbp!]
\begin{center}
  \psfrag{LI}[][][0.9]{$LI$}
  \psfrag{om}[][][0.9]{$\omega$}
  \psfrag{omm}[][][1]{$\omega$}
  \psfrag{I}[][][1]{$I$}
  \psfrag{om10}[][][1]{$\omega$}
  \psfrag{o2}[][r][1]{$\omega^{-2}$}
  \psfrag{o32}[][r][0.9]{$\omega^{-3/2}$}
  \psfrag{o12}[][r][1]{$\omega^{-1/2}$} 
  \psfrag{50}[][r][0.9]{$50$} 
  \psfrag{100}[][r][0.9]{$100$} 
  \psfrag{200}[][r][0.9]{$200$} 
  \includegraphics*[keepaspectratio, width=8cm]{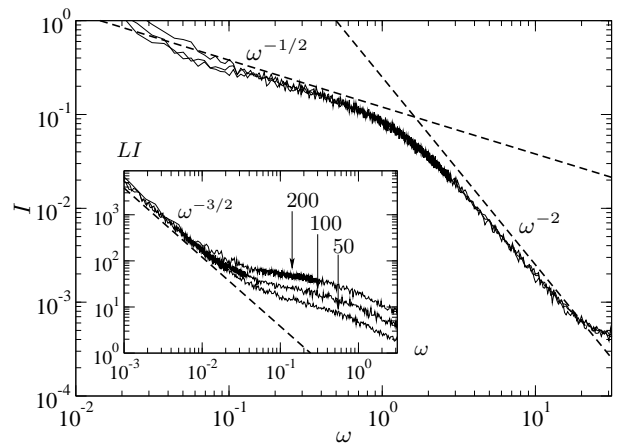}
\caption{\label{fig:spectrum0.1}
  Power spectrum for systems of $200$, $100$ and $50$ sites and rates
  $\alpha=\beta=0.1$. Averages are taken over 256 samples.  The large
  frequency behavior is dominated by local density fluctuations and
  well described within a BL theory, while the small frequency regime
  is dominated by domain wall fluctuations, as a collective mode.  The
  high resolution allows for the identification of a dynamic regime
  due to the discrete nature of density fluctuations at very short
  time. {\bf Inset:} rescaled power spectrum showing the long time
  (small frequency) regime dominated by the DW dynamics.  }
\end{center}
\end{figure}

\begin{figure}[tbp!]
\begin{center}
  \psfrag{LI}[][][0.9]{$LI$}
  \psfrag{om}[][][0.9]{$\omega$}
  \psfrag{omm}[][][1]{$\omega$}
  \psfrag{I}[][][1]{$I$}
  \psfrag{om10}[][][1]{$\omega$}
  \psfrag{o2}[][r][1]{$\omega^{-2}$}
  \psfrag{o32}[][r][0.9]{$\omega^{-3/2}$}
  \psfrag{o12}[][r][1]{$\omega^{-1/2}$} 
  \psfrag{200}[][r][0.9]{$200$} 
  \psfrag{400}[][r][0.9]{$400$} 
  \psfrag{800}[][r][0.9]{$800$} 
  \includegraphics*[keepaspectratio, width=8cm]{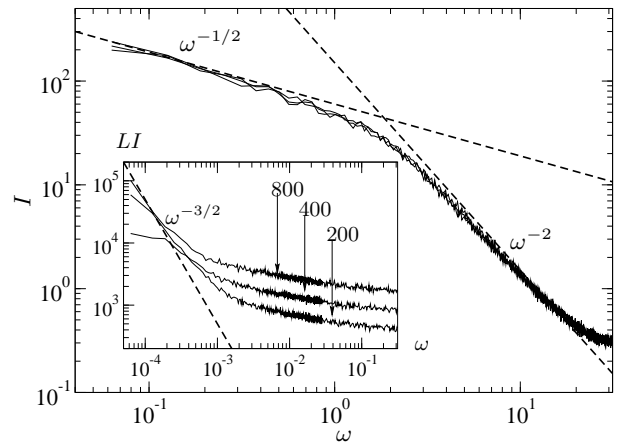}
\caption{\label{fig:spectrum0.4}
  Power spectrum for systems of $800$, $400$ and $200$ sites and rates
  $\alpha=\beta=0.4$. Averages are taken over 256 samples.  The large
  frequency behavior is dominated by local density fluctuations and
  well described within a BL theory, while the small frequency regime
  is dominated by domain wall fluctuations, as a collective mode.
  {\bf Inset:} rescaled power spectrum showing the long time regime
  dominated by the DW dynamics.  When the time scale of the DW
  and BL dynamics are comparable the separation between the two
  dynamics, although still present, is not as sharp as in
  Fig.~\ref{fig:spectrum0.1}.}
\end{center}
\end{figure}

A convenient way for visualizing the various dynamic regimes resulting
from domain wall and local density fluctuations is the power spectrum
\begin{equation}
\label{eq:spectrum}
I(\omega)\equiv\frac 1 T \avg{|\Phi(0,\omega)|^2} \, ,
\end{equation}
where $T$ is the total time of integration. It is obvious from
Figs.~\ref{fig:spectrum0.1} and~\ref{fig:spectrum0.4} that there are
three distinct dynamical regimes.

The DW theory, as described above, fully explains the low frequency
power law regime $I(\omega) \sim \omega^{-3/2}$. As can easily be
shown from specializing Eq.~(\ref{eq:coex_DW}) to $x=0$, one finds
$I(\omega) \sim L^{-1}
\omega^{-3/2}$~\cite{takesue-mitsudo-hayakawa:03}. The time window
where DW theory is valid ranges from the hopping time $\tau_1 = 1 /
\D$ to the time needed to travel a distance comparable to the system
size $\tau_L \sim L^2 / \D$ (note that $L$ is dimensionless). For
larger times one expects finite size effects. In frequency space this
corresponds to the domain $[\D/L^2,\D]$.

For frequencies larger than $\omega_1 \geq \D$ the dynamics is
dominated by local density fluctuations. Those are well described
within BL theory.  Note that contrary to the fluctuations of the
domain wall, these local density fluctuations are independent of the
system size; see Figs.~\ref{fig:spectrum0.1} and
~\ref{fig:spectrum0.4}. For time scales larger than the microscopic
hopping time of an individual particle (which we have set to $1$), one
can use the continuum version of the BL theory.  Hence for $\omega
\leq 1$ one expects $I(\omega) \sim \omega^{-1/2}$ which agrees very
well with our MC data; note that when $\alpha=\beta\approx 0$ the
distinction is clear (Fig.~\ref{fig:spectrum0.1}) while for
$\alpha=\beta\lesssim 1/2$ not only the time scales but also the
amplitudes become comparable since
$A=\alpha(1-\alpha)(1-\alpha(1-\alpha))$ and $C^{\rm {DW}}(0,0)\sim
(1-2\alpha)^2/4$ (from Eq.~\ref{eq:coex_DW}). Therefore (see
Fig.~\ref{fig:spectrum0.4}) the distinction between BL and DW regime
becomes clearly visible only at very large time (and large
systems). At the critical point the amplitude of the DW correlation is
identical to zero (see Eq.~\ref{eq:dw_sum}) and therefore the
fluctuations are described by BL in its non-linear version.

For larger frequencies one has to account for lattice effects. If one
applies the lattice version of the BL theory one finds~(see
appendix~\ref{app:bl_lattice})
\begin{eqnarray}
\label{eq:discrete_corrfou}
C_k(\omega) = \frac{2(1-\cos(\frac{k\pi}{L}))A} 
{\omega^2+ \left(1-\cos(\frac{k\pi}{L})\right)^2}\, .
\end{eqnarray}
In order to obtain the power spectrum $C_k(\omega)$ has to be summed
over all modes numbers $k$. The dominant contribution for large
frequencies are due to wave vectors close to the zone boundary, $k
=L/2$ resulting in a power spectrum $I(\omega) \sim \omega^{-2}$,
which is again well confirmed by our MC data
(Figs.~\ref{fig:spectrum0.1} and~\ref{fig:spectrum0.4}).

\section{Conclusions}

In conclusion we have analyzed the dynamics of the TASEP model over
the whole parameter range of exit and entrance rate with emphasis on
the behavior in the low and high density regime and the corresponding
phase boundary. It turns out that most of the dynamics can be nicely
explained in terms of the combined effect of local density
fluctuations and collective domain wall motion. The dynamics of the
domain wall is determined by the stochasticity in the entrance and
exit of particles at the system boundaries. Depending on the
parameters this yields to a random walk with or without drift towards
the boundaries. For the description of the local density fluctuations
we have adopted methods from kinetic theories for electronic
transport, known as Boltzmann-Langevin approach. Both, the
Boltzmann-Langevin and the domain wall approaches are to a large
extent phenomenological and hence limited in the range of
applicability. Hence we have complemented our studies by extensive
Monte-Carlo simulations of the TASEP model using the BKL algorithm
which allows us to study the real time dynamics with good accuracy.
Our main findings are as follows. For very low densities, the
linearized Boltzmann-Langevin theory accounts quantitatively for the
shape of the density-density correlations. It becomes less accurate
for densities approaching the maximal density of $1/2$ as expected
from the approximate nature of the theory. Analogous arguments apply
for very high densities by virtue of particle-hole symmetry. For
densities close to $1/2$ linearized Boltzmann-Langevin theory is
quantitatively wrong but still captures the main features
qualitatively. Exactly at the critical point, $\alpha = \beta = 1/2$,
the full Boltzmann-Langevin theory is identical to the noisy Burgers
equation which is known to be in the same universality class as the
TASEP model right at this point.

As summarized by the power spectra in Figs.~\ref{fig:spectrum0.1}
and~\ref{fig:spectrum0.4} there is a time scale separation between the
domain wall motion and the local density fluctuations. For frequencies
larger than the hopping time of the domain wall $\D$ it is the local
density fluctuations which dominate the spectrum. Upon using the
continuous and the discrete version of the linearized
Boltzmann-Langevin approach we can fully account for the crossover
from $\omega^{-1/2}$ to $\omega^{-2}$ in the spectrum. For low
frequencies $\omega < \D$ domain wall theory gives a power spectrum of
$\omega^{-3/2}$ in agreement with the Monte-Carlo data.

In summary, two rather elementary approaches, domain wall and
Boltzmann-Langevin theory, seem to capture most of the observed
dynamics of the TASEP model.  This suggests that it may be worthwhile
to look for more complex systems which also could be described by
these simple methods.

\acknowledgments We have profited from discussions with Thomas
Franosch and Jaime Santos.  We are grateful to the authors of
Ref.\cite{bhattacharjee-seno:01} for having provided the software for
numerical scaling analysis there mentioned. A.P. was supported by a
Marie-Curie Fellowship no. HPMF-CT-2002-01529 and by the Aides Jeunes
Chercheurs Universit\'e Montpellier 2. This work was supported in part
by the Junge Akademie and SFB290 (FvO).

\appendix

\section{Critical Point $\alpha = \beta = 1/2$}
\label{sec:criticality}

\begin{figure}[tbp!]
  \begin{center} \psfrag{C(x,t)}[][][1]{$C(x,t)$}
  \psfrag{Site}[][][1]{$x$}	
  \includegraphics*[keepaspectratio,width=8cm]{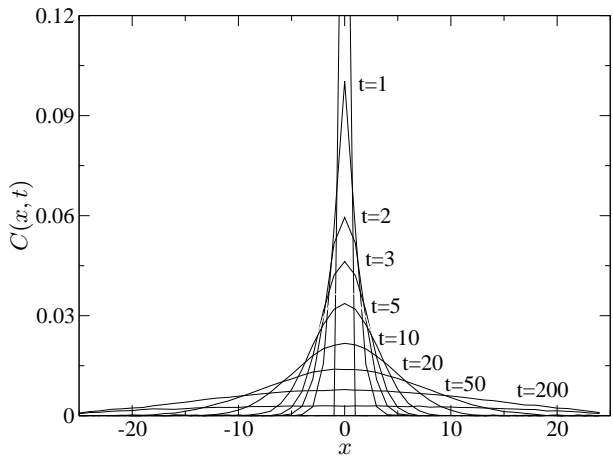} 
  \caption{\label{fig:ex4} Time series of the correlation function
$C(x,t)$ versus $x$ at the critical point ($\alpha=0.5$
  and $\beta=0.5$); the times are indicated in the graph;
 the system size is $N=50$.  Averages are taken over $10$ million samples.}
  \end{center}
\end{figure}

{F}or completeness we shortly discuss our results for the correlation
function right at the critical point ($\alpha = \beta = 1/2$). As can
be inferred from Fig.~\ref{fig:ex4} the temporal evolution of its
shape is qualitatively and quantitatively different from the low and
high density phases.

\begin{figure}[tbp!]
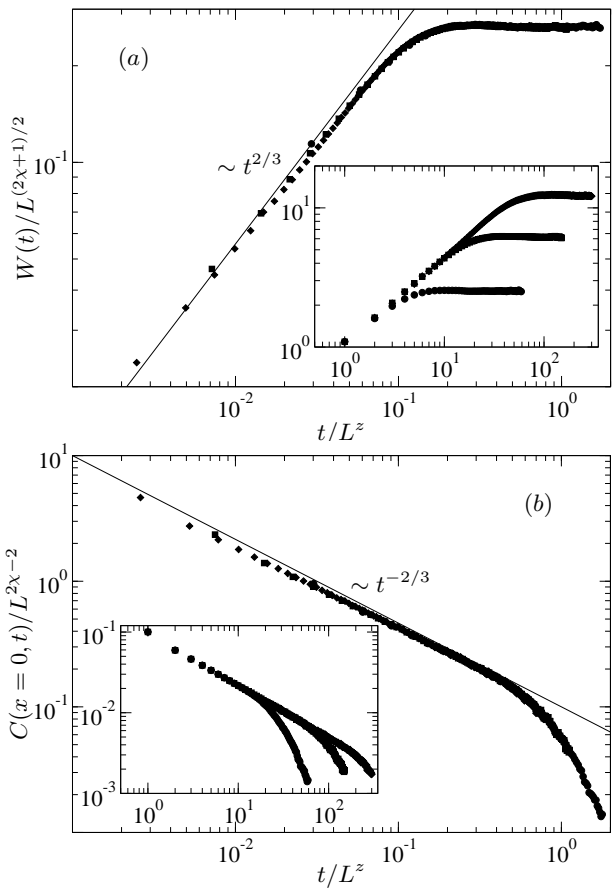

  \begin{center}
    \psfrag{C/L}[t][t][1]{$W(t)/L^{(2\chi+1)/2}$}
    \psfrag{T/L}[t][t][1]{$t/L^z$} \psfrag{x}[][][1]{$\sim t^{2/3}$}
    \psfrag{(a)}[][][1]{$(a)$} \includegraphics*[keepaspectratio,
    width=8cm]{a5b5scalwidth.eps} \vfill
    \psfrag{C/L}[][][1]{$C(x=0,t)/L^{2\chi-2}$}
    \psfrag{T/L}[][][1]{$t/L^z$} \psfrag{x}[][][1]{$\sim t^{-2/3}$}
    \psfrag{(b)}[][][1]{$(b)$} \includegraphics*[keepaspectratio,
    width=8cm]{a5b5scalpeak.eps}
    \caption{\label{fig:scal} (a) Width of the correlation function 
      and (b) autocorrelation. Averages are computed using $10^7$ samples
      at the critical point $\alpha=\beta=1/2$.  The data are rescaled
      according to Eqs.~(\ref{eq:scaling_width}) and
      (\ref{eq:scaling_autoburgers}) with the exponents presented in
      the text.  Insets show the plots before rescaling.}
  \end{center}
\end{figure}

The critical exponents are obtained from a finite size scaling
analysis of the height and width of the correlation function. The
insets of Figs.~\ref{fig:scal} a and b show Monte-Carlo data for
system sizes $N=10$, $N=25$ and $N=50$. These data can be replotted
upon using the finite size scaling relations for the height and width,
respectively,
\begin{equation}
\label{eq:scaling_autoburgers}
C(x=0,t)=L^{2\chi-2}g(t/L^z)=L^{-1}g(t/L^\frac 3 2) \, ,
\end{equation}
\begin{equation}
\label{eq:scaling_width}
W(t)= L^{\frac{2\chi+1}{2}} f(t/L^z)
    = L f(t/L^\frac 3 2) \, ,
\end{equation}
which will give us numerical values for the critical exponents.

The critical exponents were determined using an algorithm provided by
the authors of Ref.\cite{bhattacharjee-seno:01}. This code computes
and minimizes a sum which weights the distance from an interpolating
function based on all the given sequences of data. Errors are
extracted measuring the width of the minimum of such function (which
has been tested to be zero if the values are exact). From this we
obtain for the autocorrelation (peak) $2\chi=0.98\pm0.03$ and
$z=1.52\pm0.02$, and consistently $2\chi=0.92\pm 0.09$ and
$z=1.53\pm0.08$ for the width. It constitutes direct numerical
evidence for the system belonging to the KPZ universality class
($2\chi=1$ and $z=3/2$), as expected from earlier analytical results
for periodic systems.  Our measurements confirm the numerical results
in Ref.~\cite{juhasz-santen:03}.

\section{Analysis of the Boltzmann-Langevin equation on a discrete lattice}
\label{app:bl_lattice}
In this appendix we study the time behavior of the correlation
function at short times, in the regime where the discreteness of the
system plays a major role. We use $\phi_n(t)=\delta\rho_n(t)$ as the
discrete equivalent of the field $\phi(x,t)$. We can write an equation
of motion for the correlation function multiplying
Eq.~(\ref{eq:langevin_discrete}) by $\phi_n^0$:
\begin{eqnarray}
\label{eq:corr_eom}
\lefteqn{\deri{\avg{\phi_n\phi_n^0}}{t} = 
\avg{\phi_{n-1}\phi_n^0}(1-\alpha)-\avg{\phi_n\phi_n^0}+}
\nonumber\\
&+\alpha\avg{\phi_{n+1}\phi_n^0} -
\avg{\eta_n\phi_n^0}+\avg{\eta_{n-1}\phi_n^0}
\end{eqnarray}
where $\stat{\rho}=\alpha$ and we neglect the non linear terms. This
system of equations involves two point correlation functions for three
different lattice sites, but can be written in a closed form (at least
for short time regimes) by assuming that:
\begin{enumerate}
\item $\avg{\eta_n\phi_n^0}=0$ which makes sense being the noise
  independent on the dynamics itself
\item $\avg{\phi_{n-2}\phi_n^0}=0$ which is reasonable for short time
\item $\avg{\phi_{m}^0\phi_n^0}=\frac1 4\delta_{mn}$ (this originates
  the sharp tip in $C(x,t)$)
\end{enumerate}
The central site of the system will be considered the reference site
($n=0$).  Defining $C_i(t)\equiv\avg{\phi_{n+i}(t)\phi_n^0}$ we
rewrite Eq.~(\ref{eq:corr_eom}) as:
\begin{equation}
\deri{C_0}{t}=(1-\alpha)C_{-1}-C_0+\alpha C_{+1}
\end{equation}
multiplying Eq.~(\ref{eq:corr_eom} for $\phi_{n-1}$ and $\phi_{n+1}$ by
analogous reasoning we find a system of linear differential equations:
\begin{equation}
\deri{\vec{v}(t)}{t}=\hat M\vec{v}(t)
\end{equation}
with $\vec{v}=(C_{-1},C_0, C_{+1})^t$ and 
$  \hat M=
  \left[
    \begin{array}{ccc}
      -1 & \alpha & 0\\
      1-\alpha & -1 & \alpha\\
      0 & 1-\alpha & -1
    \end{array}
  \right]
$
%\end{equation}
For the initial condition $\vec v_0=(0, 1/4 , 0)^t$ the solution $\vec
v(t)=\exp (\hat M t)\vec v_0$ leads to
\begin{eqnarray}
C_{0}(t)&=&\frac 1 4 e^{-t}\cosh
\left(t\sqrt{2\alpha(1-\alpha)}\right)\nonumber\\
&=&\frac1 4 -\frac 1 4 t+\frac 1 4(\frac 1 2+\alpha-\alpha^2) t^2+O(t^3)
\end{eqnarray}
At short time the autocorrelation decays linearly in time from a
constant value ($1/4$), while the other terms grow linearly:
\begin{eqnarray}
C_{-1}(t)&=&\frac 1 4\sqrt{\frac{\alpha}{1(1-\alpha)}}e^{-t}\sinh 
\left(t\sqrt{2\alpha(1-\alpha)}\right)\nonumber\\
&=&\frac\alpha 4 t-\frac \alpha 4 t^2+O(t^3)\\
C_{+1}(t)&=&C_{+1}=\frac 1 4\sqrt{\frac{1-\alpha}{2\alpha}} 
e^{-t}\sinh  \left(t\sqrt{2\alpha(1-\alpha)}\right)\nonumber\\
&=&\frac{(1-\alpha)}{4} t-\frac{1-\alpha}{4}t^2+O(t^3)\, .
\end{eqnarray}
Note that even relaxing hypothesis ($2$), assuming therefore
$C_{\pm2}\neq0$ and dealing with a larger matrix $\hat M$, one does
not find correction to the leading behavior in time for $C_0(t)$,
since correlation functions for more distant sites, as $C_{\pm2}$,
scale as $C_{\pm2}\sim t^2$.

In order to look at the behavior in frequency space, we apply the BL
scheme in the discrete lattice and extrapolate the regime of the
correlation function at large $\omega$. Let us start from the real
space-time Boltzmann Langevin Eq.~(\ref{eq:langevin_discrete}) and let
us introduce the discrete Fourier transform
$\phi_k=\sum_{n=-L/2}^{L-1}\phi_n e^{ikn\pi/L}$ where $k$ indicate the
mode number. In order to express the linearized BL equation in
discrete Fourier space, we multiply Eq.~(\ref{eq:langevin_discrete})
by $e^{ikn\pi/L}$ and sum over $n$.  Even though the system is not
translational invariant, in this limit the system can be considered as
infinite and we do not take care of the boundaries. Performing a
Fourier transform in time we get the discrete equivalent of
Eq.~(\ref{eq:langevin}):
\begin{equation}
\left[i(\omega+2\alpha\sin \frac{k\pi}{L})+(1-e^{\frac{ik\pi}{L}})\right]\phi_k(\omega)=\eta_k(\omega)(1-e^{\frac{ik\pi}{L}})
\end{equation}
and find the correlation function
\begin{equation}
\label{eq:corr_fourier}
C_k(\omega) = \frac{2A\left(1-\cos \frac{k\pi}{L}\right)} {\left(\omega- v\sin \frac{k\pi}{L}\right)^2+\left(1-\cos \frac{k\pi}{L}\right)^2}
\end{equation}
where we use the notation $v=1-2\alpha$ and
$A=\alpha(1-\alpha(1-\alpha))$ as done above.

The dominant contribution for large frequencies are due to wave
vectors close to the zone boundary, $k =L/2$ resulting in a power
spectrum $I(\omega) \sim \omega^{-2}$, which is again nicely confirmed
by our MC data (Figs.~\ref{fig:spectrum0.1} and~\ref{fig:spectrum0.4}).

The autocorrelation is the sum over all the modes, but the dominant
contribution for large frequencies are due to wave vectors close to
the zone boundary, $k =L/2$:
\begin{equation}
C(x=0,\omega)\simeq\frac1L\sum_{k=0}^{L-1}\frac{2A\left(1-\cos \frac{k\pi}{L}\right)}{\omega^2}=2A\omega^{-2}
\end{equation}
which (by Wiener-Khinchin theorem) is the power spectrum mentioned in
Sec.\ref{sub:coex}.

\end{document}